\documentclass[11pt]{article}
\usepackage{epsfig} 
\newcommand{\sect}[1]{ \section{#1} \setcounter{equation}{0} }

\newcommand{\Dslash}{D \! \! \! \! /} 
 
\newcommand{\pslash}{p \! \! \! /}

\newcommand{\half}{\mbox{\small{$\frac{1}{2}$}}} 
\newcommand{\MSbar}{\overline{\mbox{MS}}} 
 
\newcommand{\Nf}{N_{\!f}}
\newcommand{\NF}{N_{\!F}}
\newcommand{\NA}{N_{\!A}}

\begin{document}
\title{Two loop $\MSbar$ Gribov mass gap equation with massive quarks}
\author{F.R. Ford \& J.A. Gracey, \\ Theoretical Physics Division, \\ 
Department of Mathematical Sciences, \\ University of Liverpool, \\ P.O. Box 
147, \\ Liverpool, \\ L69 3BX, \\ United Kingdom.} 
\date{} 
\maketitle 
\vspace{5cm} 
\noindent 
{\bf Abstract.} We compute the two loop $\MSbar$ correction to the Gribov mass 
gap equation in the Landau gauge using the Gribov-Zwanziger Lagrangian with 
massive quarks included. The computation involves dilogarithms of complex 
arguments and reproduces the known gap equation when the quark mass tends to 
zero.  

\vspace{-15cm}
\hspace{13.5cm}
{\bf LTH 831}

\newpage 

\sect{Introduction.} 

The quantum field theory underlying the strong nuclear force is Quantum
Chromodynamics (QCD). It is an extension of Quantum Electrodynamics (QED) where
the gauge fields are required to be elements of a non-abelian colour group,
$SU(3)$, as opposed to the abelian $U(1)$ of electric charge. Whilst this is a
simple mathematical generalization the properties of the Yang-Mills field
theory are significantly different. Clearly QCD is asymptotically free which is
not unrelated to the fact that the basic fields analogous to electrons 
correspond to particles which are never isolated in nature, called quarks. They
are held together in pairs or triplets by the quanta of the strong force called
gluons. Equally these have never been seen isolated in experiments but rather 
at high energy they are effectively massless asymptotically free fields which 
to all intents and purposes behave as massless fundamental particles. To a 
degree this behaviour is parallel to the properties of the photons and 
electrons of QED. However, both fundamental forces differ in behaviour in the 
infrared region. For instance, in QCD infrared slavery dominates the 
confinement picture and the gluon propagator does not have the behaviour of a 
massless fundamental particle. One situation where this property can be 
manifestly seen is in Gribov's construction of the gluon propagator at low 
energy in the Landau gauge, \cite{1}. An additional divergence in the structure
of QED and QCD emanates from the way one tries to fix a (linear) covariant 
gauge. In QED one can fix the gauge in a global sense. By contrast, Gribov 
pointed out, \cite{1}, that in Yang-Mills theory the covariant gauge condition 
for the Landau gauge has an ambiguity. This occurs at zeroes of the 
Faddeev-Popov operator when different gauge configurations satisfy the 
{\em same} gauge fixing condition. In a local region in the neighbourhood of 
the origin of configuration space, where perturbation theory is valid, there is
no such ambiguity and standard perturbative calculations are perfectly adequate
to describe ultraviolet behaviour. However, to properly fix the gauge globally 
the problem of Gribov copies must be taken into account in defining the path 
integral of the theory, \cite{1}. Gribov achieved this by restricting the path 
integral to the region of configuration space containing the first Gribov 
region, denoted by $\Omega$. This is defined to be the region containing the 
origin where the Faddeev-Popov operator, 
${\cal M}(A)$~$\equiv$~$-$~$\partial^\mu D_\mu(A)$, is strictly positive. 
Consequently, the path integral is cutoff and a natural mass parameter,
$\gamma$, called the Gribov mass emerges, \cite{1}. It is not an independent 
parameter of Yang-Mills but is non-perturbative and satisfies a gap equation. 
In turn this gap equation derives from the restriction of the path integral to 
$\Omega$ by the no pole condition, \cite{1}. In other words the average of 
$1/{\cal M}(A)$ over $\Omega$ is finite. This construction radically alters
the infrared properties of the theory. For instance, it leads to a gluon 
propagator which is not fundamental in the sense that it has no (real) pole, 
\cite{1}.  Moreover, it is suppressed in the infrared since it vanishes in the 
infrared limit. Further, the gap equation implies that the propagator of the 
Faddeev-Popov ghost is not fundamental but has a dipole behaviour at low 
momenta which is referred to as ghost enhancement. These infrared properties of
the constituent fields are believed to be related to confinement, \cite{1}, and
over the years has led to intense interest in studying gluon and ghost 
$2$-point functions on the lattice and with Dyson Schwinger equation (DSE) 
methods.  
 
Another approach was also developed, however, in a series of articles by 
Zwanziger and collaborators, \cite{2,3,4,5,6,7,8,9,10}, with other relevant 
contributions in, for instance, \cite{11,12}. In essence the semi-classical 
approach of Gribov for Landau gauge Yang-Mills was put on a firmer footing with
the construction of a localized renormalizable Lagrangian, \cite{3,4,7,8}. The 
renormalizability being established by various authors, \cite{8,13,14}. The 
implementation of the horizon condition defining $\Omega$ in the original 
approach led to a non-local operator in the action which clearly inhibits 
direct calculations. In \cite{3,4,7,8} Zwanziger localized the non-locality 
with a (finite) set of extra fields which defined the horizon condition in an 
equivalent fashion. The beauty of the renormalizability, \cite{8,13,14}, aside 
from allowing for calculations was to demonstrate that {\em none} of the known 
and accepted properties of QCD at high energy were changed or upset. For 
instance, asymptotic freedom remains with the {\em same} $\beta$-function. 
However, the advantage of the new formulation was to allow for loop 
calculations and the extension to the next level of computation of the gap 
equation, gluon suppression and ghost enhancement. This was achieved in 
\cite{15} and \cite{16}. In the former the two loop $\MSbar$ gap equation for 
$\gamma$ was established when massless quarks are present. This was checked in 
a non-trivial way by verifying that ghost enhancement was satisfied at two 
loops precisely when $\gamma$ obeyed the gap equation. Indeed the theory has no
meaning as a gauge theory unless $\gamma$ does this and hence is not an 
independent parameter of the theory, \cite{1}. In the latter article, 
\cite{16}, the one loop gluon suppression was verified as well as the 
{\em exact} evaluation of all the one loop $2$-point functions of the fields of
the Gribov-Zwanziger Lagrangian.

Given this background we come to the main purpose of this article. Clearly in
the real world quarks are not massless but massive. Therefore, to have a more
realistic understanding of the Gribov situation it seems appropriate to include
massive quarks. As will be evident from what is recorded here this is far from 
a trivial task. First, quarks only appear diagrammatically in the gap equation 
first at two loops. Moreover, this results in Feynman integrals involving three
scales. Aside from the quark mass itself, the gluon propagator actually has two
mass scales in the sense of a conventional fundamental propagator. These are 
$\pm i \sqrt{C_A} \gamma^2$ where the mass is actually imaginary. (The presence
of $\sqrt{C_A}$ stems from our conventions which follow those derived in 
\cite{15,16}.) The presence of the imaginary mass further complicates Feynman 
integral evaluation since some of the fundamental functions of one and two loop 
integrals, such as dilogarithms, need to be considered for complex arguments.
Therefore, it is the main purpose of this article to extend the massless quark
two loop $\MSbar$ gap equation of \cite{15} to the massive quark case. 
Moreover, we will discuss the effect it has on the enhancement of the
Faddeev-Popov ghost. Finally, we note that given recent developments concerning
the scaling versus decoupling solutions, \cite{17,18,19,20,21,22}, for which 
there has yet to be a definitive resolution, we note that our computations will
be the foundation for extensions to the decoupling gap equation. This will be 
required if that solution is eventually established as the correct picture. 
Moreover, this is possible in our approach because the decoupling solution can 
be accommodated in the Gribov-Zwanziger formulation, \cite{23,24}. Though it 
will in fact be a more difficult task than the current work due to the 
generation of mass for the localizing Zwanziger ghost fields. 

The paper is organised as follows. Section two is devoted to reviewing the
relevant aspects of the Gribov-Zwanziger formalism for the massive quark two
loop gap equation. The construction of the two loop scalar master integrals to
the finite part is presented in section three where we discuss at length their
expression in terms of functions of real variables. This is necessary in order 
to produce a real gap equation rather than a form which has functions of
complex variables due to the gluon widths. Our main result is provided in 
section four whilst we draw our conclusions in section five. 

\sect{Formalism.}

In this section we recall the relevant aspects of the basic Gribov-Zwanziger
Lagrangian we will use to extend the results of \cite{15}. From \cite{3,4,7,8}
the (bare) Lagrangian is  
\noindent
\begin{eqnarray}
L^{\mbox{\footnotesize{GZ}}} &=& L^{\mbox{\footnotesize{QCD}}} ~+~ 
\bar{\phi}^{ab \, \mu} \partial^\nu \left( D_\nu \phi_\mu \right)^{ab} ~-~ 
\bar{\omega}^{ab \, \mu} \partial^\nu \left( D_\nu \omega_\mu \right)^{ab} 
\nonumber \\  
&& -~ g f^{abc} \partial^\nu \bar{\omega}_\mu^{ae} \left( D_\nu c \right)^b
\phi^{ec \, \mu} ~+~ \frac{\gamma^2}{\sqrt{2}} \left( f^{abc} A^{a \, \mu} 
\phi^{bc}_\mu ~-~ f^{abc} A^{a \, \mu} \bar{\phi}^{bc}_\mu \right) ~-~ 
\frac{d \NA \gamma^4}{2g^2} 
\label{laggz}
\end{eqnarray} 
where we use the usual linear covariant gauge fixing prescription
\begin{equation} 
L^{\mbox{\footnotesize{QCD}}} ~=~ -~ \frac{1}{4} G_{\mu\nu}^a 
G^{a \, \mu\nu} ~-~ \frac{1}{2\alpha} (\partial^\mu A^a_\mu)^2 ~-~ 
\bar{c}^a \partial^\mu D_\mu c^a ~+~ i \bar{\psi}^{iI} \Dslash \psi^{iI} ~-~
m_q \bar{\psi}^{iI} \psi^{iI} ~. 
\label{lagqcd}
\end{equation} 
Although we will work strictly in the Landau gauge we have included the usual
gauge fixing parameter $\alpha$ since it is required to derive the gluon
propagator. Aside from this it should be understood that $\alpha$ is set to
zero throughout. Briefly our conventions in (\ref{laggz}) and (\ref{lagqcd}) 
are that $A^a_\mu$ is the gluon, $c^a$ is the Faddeev-Popov ghost, $\psi^{iI}$
is the quark with mass $m_q$ and $\phi^{ab}_\mu$, $\bar{\phi}^{ab}_\mu$, 
$\omega^{ab}_\mu$ and $\bar{\omega}^{ab}_\mu$ are the Zwanziger localizing 
ghosts. The latter pair are anti-commuting like the Faddeev-Popov ghosts 
whereas $\phi^{ab}_\mu$ and $\bar{\phi}^{ab}_\mu$ are commuting. The Lagrangian
is expressed in $d$-dimensional spacetime since we will use dimensional 
regularization throughout to isolate the divergence structure of the Feynman 
graphs where $d$~$=$~$4$~$-$~$2\epsilon$ and $\epsilon$ is the regularizing 
parameter. The various indices have the ranges $1$~$\leq$~$I$~$\leq$~$\Nf$,
$1$~$\leq$~$a$~$\leq$~$\NA$ and $1$~$\leq$~$i$~$\leq$~$\NF$ where $\Nf$ is the 
number of quark flavours and $\NF$ and $\NA$ are the respective dimensions of 
the fundamental and adjoint representations. The various covariant derivatives
are
\begin{eqnarray}
D_\mu c^a &=& \partial_\mu c^a ~-~ g f^{abc} A^b_\mu c^c \nonumber \\
D_\mu \psi^{iI} &=& \partial_\mu \psi^{iI} ~+~ i g T^a_{IJ} A^a_\mu \psi^{iJ}
\nonumber \\
\left( D_\mu \phi_\nu \right)^{ab} &=& \partial_\mu \phi^{ab}_\nu ~-~
g f^{acd} A^c_\mu \phi^{db}_\nu 
\end{eqnarray}
where $g$ is the coupling constant, $G^a_{\mu\nu}$ is the usual gluon field
strength and $T^a$ are the generators of the colour group which has structure
functions $f^{abc}$. We note that we have reverted to the conventions of the
original form of the Lagrangian, \cite{4,8}, in the mixed $2$-point 
sector\footnote{In \cite{15,16} the Feynman rules of this Lagrangian were used 
within the computer algebra computations though the actual Lagrangian recorded 
in the articles followed the conventions of \cite{14}.}. With this formulation 
of the Gribov-Zwanziger Lagrangian we have checked that the results of the 
massless quark gap equation at two loops and Faddeev-Popov ghost enhancement 
correctly emerge. The fields $\phi^{ab}_\mu$ and $\bar{\phi}^{ab}_\mu$ 
correspond to the localization of the Gribov horizon condition which originally
was
\begin{equation}
\left\langle A^a_\mu(x) \frac{1}{\partial^\nu D_\nu} A^{a\,\mu}(x)
\right\rangle ~=~ \frac{d N_A}{C_A g^2}
\end{equation}
and now equates to
\begin{eqnarray}
f^{abc} \langle A^{a \, \mu}(x) \phi^{bc}_\mu(x) \rangle &=& 
\frac{d \NA \gamma^2}{\sqrt{2}g^2} \nonumber \\
f^{abc} \langle A^{a \, \mu}(x) \bar{\phi}^{bc}_\mu(x) \rangle &=& -~ 
\frac{d \NA \gamma^2}{\sqrt{2}g^2} ~. 
\label{gapdef}
\end{eqnarray} 
Our conventions are actually crucial to reproducing the correct form of the 
gluon propagator of the original Gribov article, \cite{1}. Using other 
conventions could lead to, for example, a gluon propagator which has a normal 
mass as well as a tachyonic mass. From (\ref{laggz}) and (\ref{lagqcd}) we have
checked that the propagators of the fields, with momentum $p$, are 
\begin{eqnarray}
\langle A^a_\mu(p) A^b_\nu(-p) \rangle &=& -~ 
\frac{\delta^{ab}p^2}{[(p^2)^2+C_A\gamma^4]} P_{\mu\nu}(p) \nonumber \\  
\langle A^a_\mu(p) \bar{\phi}^{bc}_\nu(-p) \rangle &=& -~ 
\frac{f^{abc}\gamma^2}{\sqrt{2}[(p^2)^2+C_A\gamma^4]} P_{\mu\nu}(p) 
\nonumber \\  
\langle \phi^{ab}_\mu(p) \bar{\phi}^{cd}_\nu(-p) \rangle &=& -~ 
\frac{\delta^{ac}\delta^{bd}}{p^2}\eta_{\mu\nu} ~+~  
\frac{f^{abe}f^{cde}\gamma^4}{p^2[(p^2)^2+C_A\gamma^4]} P_{\mu\nu}(p) 
\nonumber \\ 
\langle \omega^{ab}_\mu(p) \bar{\omega}^{cd}_\nu(-p) \rangle &=& -~ 
\frac{\delta^{ac}\delta^{bd}}{p^2} \eta_{\mu\nu} \nonumber \\ 
\langle c^a(p) \bar{c}^b(-p) \rangle &=& \frac{\delta^{ab}}{p^2} \nonumber \\ 
\langle \psi^{iI}(p) \bar{\psi}^{jJ}(-p) \rangle &=& 
\delta^{ij}\delta^{IJ} \frac{(\pslash + m_q)}{[p^2+m_q^2]} 
\label{props}
\end{eqnarray} 
in the Landau gauge where 
\begin{equation}
P_{\mu\nu}(p) ~=~ \eta_{\mu\nu} ~-~ \frac{p_\mu p_\nu}{p^2} 
\end{equation}
is the usual projector. We have retained a non-zero $\alpha$ in inverting the
matrix of $2$-point functions in the quadratic part of the Lagrangian before
setting $\alpha$~$=$~$0$ to recover the Landau gauge. The Feynman rules for
the vertices have no convention complications and are straightforward to
derive from (\ref{laggz}) and (\ref{lagqcd}). Though we note that the explicit
cubic interaction of (\ref{laggz}) is completely passive since it is never
present within the Feynman diagrams contributing to any Green's function of
interest at the two loop level of this article. 

As (\ref{laggz}) is renormalizable and incorporates the Gribov properties we
now discuss the set-up for our computation. The gap equation satisfied by
$\gamma$ is defined by the no-pole condition determining the boundary of
$\Omega$, \cite{1}. In the original approach of \cite{1} this equated to 
evaluating the vacuum expectation value of $f^{abc} A^{a\,\mu} \phi^{ab}_\mu$ 
and ensuring it satisfied (\ref{gapdef}) where the right side is a finite 
object, \cite{3,4,7,8}. However, at this point we note that in all the vacuum 
expectation values one has to take into account the renormalization of the 
fields and parameters. In this respect we note that the anomalous dimensions of
all the quantities we require are available at three loops in the $\MSbar$ 
scheme for an arbitrary colour group, \cite{13,14,15,16}. Further, at four 
loops the renormalization of $\gamma$ is known for the $SU(N_c)$ Lie colour 
groups, \cite{25}. These follow partly through the renormalizability of 
(\ref{laggz}), \cite{8,13,14}, but also because the localizing fields and 
$\gamma$ do not undergo independent renormalization in the Landau gauge. 
Instead all the renormalization constants are determined by Slavnov-Taylor 
identities, \cite{8,13,14}. Denoting the associated anomalous dimensions of a 
field or parameter $\Gamma$ by $\gamma_\Gamma(a)$ where $a$~$=$~$g^2/(16\pi^2)$
then we record the renormalization constants we require as being encoded in the 
anomalous dimensions  
\begin{eqnarray} 
\gamma_A(a) &=& \left[ 8 T_F \Nf - 13 C_A \right] \frac{a}{6} ~+~ \left[ 
40 C_A T_F \Nf + 32 C_F T_F \Nf - 59 C_A^2 \right] \frac{a^2}{8} ~+~ O(a^3)
\nonumber \\
\gamma_\phi(a) &=& \gamma_\omega(a) ~=~  -~ \frac{3}{4} C_A a ~+~ \left[ 
40 C_A T_F \Nf - 95 C_A^2 \right] \frac{a^2}{48} ~+~ O(a^3) \nonumber \\  
\gamma_\gamma(a) &=& \left[ 16 T_F \Nf - 35 C_A \right] \frac{a}{48} ~+~ 
\left[ 280 C_A T_F \Nf - 449 C_A^2 + 192 C_F T_F \Nf \right]
\frac{a^2}{192} ~+~ O(a^3) \nonumber \\  
\end{eqnarray} 
with the $\beta$-function 
\begin{equation} 
\beta(a) ~=~ -~ \left[ \frac{11}{3} C_A - \frac{4}{3} T_F \Nf \right] a^2 ~-~
\left[ \frac{34}{3} C_A^2 - 4 C_F T_F \Nf - \frac{20}{3} C_A T_F \Nf \right]
a^3 ~+~ O(a^4) ~.
\end{equation} 
The elementary group Casimirs are defined by
\begin{equation}
\mbox{Tr} \left( T^a T^b \right) ~=~ T_F \delta^{ab} ~~,~~
T^a T^a ~=~ C_F I ~~,~~ f^{acd} f^{bcd} ~=~ C_A \delta^{ab} ~.
\end{equation}
Whilst the higher order expressions are available we only provide them at two 
loops as that is the order we compute to here. In (\ref{gapdef}) we note that 
the renormalization of all fields and parameters present is therefore already 
fixed and hence after all contributing Feynman diagrams have been computed and 
assembled the resulting vacuum expectation value is finite. With a massive 
quark present, its mass will be renormalized in principle too. However, as it 
first appears at two loops, scaling it from a bare to a renormalized parameter 
will not affect the two loop gap calculation as the counterterms from the quark
mass renormalization constant will only arise first at three loops. 

As (\ref{gapdef}) is the vacuum expectation value of two fields it is easy to
determine since essentially it is the closure of the legs on the mixed 
propagator of (\ref{props}) and integrated over the momentum $p$. Thus for 
higher loop calculations one simply evaluates the relevant Feynman diagrams 
which are merely vacuum bubbles with various configurations of masses. We 
devolve to a later section the more detailed structure of such two loop massive
vacuum bubbles and concentrate in the remainder of this section on more general 
aspects of the two loop gap equation calculation. The main ingredients are the
generation of the Feynman graphs via the {\sc Qgraf} package, \cite{26}, and
its conversion into the symbolic manipulation language {\sc Form}, \cite{27}. 
We use {\sc Form} as it is ideal for handling the underlying algebra in an
efficient manner. For the gap equation, due to the mixed propagators, there
are $1$ one loop and $17$ two loop Feynman diagrams to be determined exactly
as a function of $\gamma$ and $m_q$. As they resolve into the basic structure
of two loop vacuum bubbles, we note that to make contact with known results we 
apply elementary partial fractions to the common factor in the propagators of 
the Gribov related fields, such as 
\begin{equation}
\frac{p^2}{[(p^2)^2 + C_A\gamma^4]} ~=~ \frac{1}{2} \left(
\frac{1}{[p^2+i\sqrt{C_A}\gamma^2]} ~+~ \frac{1}{[p^2-i\sqrt{C_A}\gamma^2]}
\right) ~.
\end{equation}
Moreover, within our {\sc Form} routines the Feynman rules are substituted
automatically and the elementary group theory is evaluated making extensive use
of the Jacobi identity for the structure functions, partly due to the form of 
the pure $\phi^{ab}_\mu$ propagator. Therefore, all that remains in determining
the gap equation for massive quarks is the substitution of the explicit forms 
for the master Feynman integrals which the {\sc Form} routines produce. The 
next section is devoted to this where we concentrate on the intricacies of 
dealing with vaccum integrals with massive quarks and complex gluon masses.  

\sect{Master integrals.}

There are two main scalar master integrals which arise in the computation. The
first is the massive one loop vacuum bubble which is virtually trivial in
comparison with that we have to consider at two loops. Though it does arise in
the two loop computation when a line in the basic form of a two loop vacuum
bubble graph is omitted. Therefore, defining 
\begin{equation}
I_1(m^2)~=~\int_k \frac {1} {\left[k^2 + m^2\right]} 
\end{equation}
where
\begin{equation}
\int_k ~\equiv~ \int \frac{d^d k}{(2\pi)^d}
\end{equation}
includes the $d$-dimensional momentum space measure we have exactly
\begin{equation}
I_1(m^2) ~=~ \frac{\Gamma(1-\half d)}{(4\pi)^{d/2}} ( m^2 )^{\half d-1} 
\end{equation}
which is trivial to expand in powers of $\epsilon$. Therefore, we now 
concentrate on the basic massive scalar two loop vacuum bubble which we define 
as 
\begin{equation}
I_2(m_x^2,m_y^2,m_z^2)~=~\int_k \int_l \frac {1}
{\left[k^2 + m^2_x\right]\left[(k-l)^2+m^2_y\right]\left[l^2+m^2_z\right]} ~. 
\end{equation}
which is completely symmetric in its arguments and has been studied extensively
over the years. See, for example, \cite{28,29,30}. Its expansion in powers of 
$\epsilon$, where $d$~$=$~$4$~$-$~$2\epsilon$, is known to several orders but 
for our purposes it suffices to record it to the finite part. For this we 
follow the notation and conventions of \cite{29}. Then we have 
\begin{eqnarray}
(4\pi)^4 I_2(x,y,z) &=& -~ \frac{c}{2\epsilon^2} ~-~ \frac{1}{\epsilon} 
\left[ \frac{3c}{2} - \mbox{L}_1 \right] \nonumber \\
&& -~ \frac{1}{2} \left[ L_2 - 6 L_1 ~+~ \xi(x,y,z) ~+~ c 
\left(7+\zeta(2)\right) \right. \nonumber \\
&& \left. ~~~~~~+~ \left(y+z-x\right) \overline{\ln}(y) \overline{\ln}(z) ~+~ 
\left(z+x-y\right) \overline{\ln}(z) \overline{\ln}(y) \right. \nonumber \\
&& \left. ~~~~~~+~ \left(y+x-z\right) \overline{\ln}(y) \overline{\ln}(x) 
\right] ~+~ O(\epsilon) 
\label{i2def}
\end{eqnarray}
where we define  
\begin{eqnarray} 
L_i &=& x\overline{\ln}^i(x) + y\overline{\ln}^i(y) + z\overline{\ln}^i(z) 
\nonumber \\ 
c &=& x ~+~ y ~+~ z \nonumber \\
a &=& \frac{1}{2} \left[ x^2 + y^2 + z^2 - 2xy - 2xz - 2yz \right]^{1/2} ~. 
\end{eqnarray} 
We also use the same notation as \cite{29} in defining
\begin{equation}
\overline{\ln} (m^2) ~=~ \ln \left( \frac{m^2}{\mu^2} \right)
\end{equation}
where $\mu$ is the mass scale which enters when using dimensional 
regularization to ensure the coupling constant remains dimensionless in
$d$-dimensions. The key part of this $\epsilon$ expansion is the function
$\xi(x,y,z)$ whose explicit form depends on the sign of the combination of
masses denoted by $a^2$. For our purposes we note that for $a^2$~$>$~$0$ then,
\cite{29},
\begin{equation}
\xi(x,y,z) ~=~ 8a \left[ M(\phi_z) + M(\phi_y) - M(-\phi_x) \right] 
\end{equation}
where 
\begin{equation} 
M(\phi) ~=~ - \int_0^\phi d \theta \ln(\sinh(\theta)) 
\end{equation}
and
\begin{equation}
\phi_x ~=~ \coth^{-1} \left[ \frac{c-2x}{2a} \right] ~. 
\end{equation} 
Moreover, we note that
\begin{equation}
\coth^{-1}(z) ~=~ \frac{1}{2}\ln\left[\frac{z+1}{z-1}\right] 
\end{equation}
and the integral defined by the intermediate function $M(\phi)$ can be written
in terms of known functions  
\begin{equation}
M(\phi) ~=~ \phi\ln(2) ~-~ \half \phi^2 ~+~ \half \zeta(2) ~-~ 
\mbox{Li}_2(e^{-\phi}) ~-~ \mbox{Li}_2(-e^{-\phi}) 
\end{equation}
where $\mbox{Li}_2(z)$ is the dilogarithm function, \cite{31},
\begin{equation}
\mbox{Li}_2(z) ~=~ -~ \int_0^z \frac{\ln(1-x)}{x} dx
\label{dilog}  
\end{equation}
and $\zeta(z)$ is the Riemann zeta function. As an exercise to aid the 
interested reader it is instructive to consider the elementary case 
$I_2(0,0,m^2)$ which is explicitly  
\begin{equation}
I_2(0,0,m^2) ~=~ \int_k \int_l \frac {1}{k^2 (k-l)^2 \left[l^2+m^2\right]} ~.
\end{equation}
It can be evaluated directly and then compared with (\ref{i2def}) to give
\begin{equation}
I_2(0,0,m^2) ~=~ -~ \frac{m^2}{2\epsilon^2} ~-~ \frac{m^2}{2\epsilon} 
\left[ 3 - 2 \overline{\ln}(m^2) \right] 
~-~ \frac{m^2}{2} \left[ 7 + 3 \zeta(2) + 2\overline{\ln}^2(m^2) 
- 6 \overline{\ln}(m^2) \right] ~+~ O(\epsilon) 
\label{int01}
\end{equation} 
which will be required for checking our expressions in the massless quark 
limit. 

However, as we are ultimately interested in the massive quark case, we have to
consider several master integrals. These are 
\begin{eqnarray}
I_2(m_q^2,m_q^2,i\sqrt{C_A}\gamma^2) &=& \int_k \int_l \frac {1}
{\left[k^2 + m^2_q\right]\left[(k-l)^2+m^2_q\right]\left[l^2+i\sqrt{C_A} 
\gamma^2\right]} \nonumber \\
I_2(m_q^2,m_q^2,-i\sqrt{C_A}\gamma^2) &=& \int_k \int_l \frac {1}
{\left[k^2 + m^2_q\right]\left[(k-l)^2+m^2_q\right]\left[l^2-i\sqrt{C_A} 
\gamma^2\right]} 
\label{mas1}
\end{eqnarray}
and the related integrals
\begin{eqnarray}
\bar{I}_2(m_q^2,m_q^2,i\sqrt{C_A}\gamma^2) &=& \int_k \int_l \frac {1}
{\left[k^2 + m^2_q\right]\left[(k-l)^2+m^2_q\right]\left[l^2+i\sqrt{C_A} 
\gamma^2\right]^2} \nonumber \\
\bar{I}_2(m_q^2,m_q^2,-i\sqrt{C_A}\gamma^2) &=& \int_k \int_l \frac {1}
{\left[k^2 + m^2_q\right]\left[(k-l)^2+m^2_q\right]\left[l^2-i\sqrt{C_A} 
\gamma^2\right]^2} ~.
\label{mas2}
\end{eqnarray}
We concentrate on the former two as the definition of the latter follows from
using elementary calculus. For the first we will focus on 
\begin{eqnarray} 
\xi(i\sqrt{C_A}\gamma^2,m^2_q,m^2_q) &=& 8a \left[ 
2M(\phi_{m^2_q}) - M(-\phi_{i\sqrt{C_A}\gamma^2}) \right] 
\end{eqnarray} 
where now 
\begin{equation}
a ~=~ \frac{i}{2} \sqrt{C_A\gamma^4 + 4i\sqrt{C_A}\gamma^2m^2_q} ~~~,~~~
c ~=~ i\sqrt{C_A}\gamma^2 ~+~ 2m^2_q ~. 
\end{equation} 
leading to the intermediate variables 
\begin{eqnarray}
\phi_{i\sqrt{C_A}\gamma^2} &=& \coth^{-1} \left[ \frac
{-\sqrt{C_A}\gamma^2-2im^2_q}{\sqrt{C_A\gamma^4+4i\sqrt{C_A}\gamma^2m^2_q}}
\right] \nonumber \\
\phi_{m^2_q} &=& \coth^{-1} \left[ \frac{\sqrt{C_A}\gamma^2}
{\sqrt{C_A\gamma^4+4i\sqrt{C_A}\gamma^2m^2_q}} \right] ~. 
\end{eqnarray} 
Though for practical purposes it is more appropriate to re-express these by
applying the logarithm definition
\begin{eqnarray}
e^{-\phi_{i\sqrt{C_A}\gamma^2}} &=& \frac
{\sqrt{C_A}\gamma^2 + \sqrt{C_A\gamma^4+4i\sqrt{C_A}\gamma^2m^2_q}}       
{\sqrt{C_A}\gamma^2 - \sqrt{C_A\gamma^4+4i\sqrt{C_A}\gamma^2m^2_q}} 
\nonumber \\  
e^{-\phi_{m^2_q}} &=& \sqrt{\frac
{\sqrt{C_A}\gamma^2 - \sqrt{C_A\gamma^4+4i\sqrt{C_A}\gamma^2m^2_q}}       
{\sqrt{C_A}\gamma^2 + \sqrt{C_A\gamma^4+4i\sqrt{C_A}\gamma^2m^2_q}}} ~.
\end{eqnarray}  
These naturally lead to the two functions 
\begin{eqnarray} 
M(t_{m^2_q}) &=& \frac{\zeta(2)}{2} ~+~ \frac{1}{2} \ln \left[ \frac
{\sqrt{C_A}\gamma^2 +\sqrt{C_A\gamma^4+4i\sqrt{C_A}\gamma^2m^2_q}}
{\sqrt{C_A}\gamma^2 -\sqrt{C_A\gamma^4+4i\sqrt{C_A}\gamma^2m^2_q}} \right]
\ln(2) \nonumber \\
&& -~ \frac{1}{8} \ln^2 \left[ \frac
{\sqrt{C_A}\gamma^2 +\sqrt{C_A\gamma^4+4i\sqrt{C_A}\gamma^2m^2_q}}
{\sqrt{C_A}\gamma^2 -\sqrt{C_A\gamma^4+4i\sqrt{C_A}\gamma^2m^2_q}} \right] 
\nonumber \\ 
&& -~ \frac{1}{2}\mbox{Li}_2\left[ \frac
{\sqrt{C_A}\gamma^2 - \sqrt{C_A\gamma^4+4i\sqrt{C_A}\gamma^2m^2_q}}       
{\sqrt{C_A}\gamma^2 + \sqrt{C_A\gamma^4+4i\sqrt{C_A}\gamma^2m^2_q}} \right] 
\end{eqnarray} 
and 
\begin{eqnarray}
M(-\phi_{i\sqrt{C_A}\gamma^2}) &=& \frac{\zeta(2)}{2} ~+~
\ln \left[ \frac{\sqrt{C_A}\gamma^2 
+\sqrt{C_A\gamma^4+4i\sqrt{C_A}\gamma^2m^2_q}}  
{\sqrt{C_A}\gamma^2 -\sqrt{C_A\gamma^4+4i\sqrt{C_A}\gamma^2m^2_q}}
\right]\ln(2) \nonumber \\  
&& -~ \frac{1}{2} \ln^2 \left[ \frac
{\sqrt{C_A}\gamma^2 +\sqrt{C_A\gamma^4+4i\sqrt{C_A}\gamma^2m^2_q}}  
{\sqrt{C_A}\gamma^2 -\sqrt{C_A\gamma^4+4i\sqrt{C_A}\gamma^2m^2_q}}
\right] \nonumber \\
&& -~ \mbox{Li}_2\left[ \frac
{\sqrt{C_A}\gamma^2 - \sqrt{C_A\gamma^4+4i\sqrt{C_A}\gamma^2m^2_q}}        
{\sqrt{C_A}\gamma^2 + \sqrt{C_A\gamma^4+4i\sqrt{C_A}\gamma^2m^2_q}} \right]
\nonumber \\
&& -~ \mbox{Li}_2\left[ \frac
{\sqrt{C_A\gamma^4+4i\sqrt{C_A}\gamma^2m^2_q} - \sqrt{C_A}\gamma^2}        
{\sqrt{C_A}\gamma^2 + \sqrt{C_A\gamma^4+4i\sqrt{C_A}\gamma^2m^2_q}} \right] 
\end{eqnarray} 
where we have used the relationship, \cite{31}, 
\begin{eqnarray}
\mbox{Li}_2(x) ~+~ \mbox{Li}_2(-x) ~=~ \frac{1}{2} \mbox{Li}_2(x^2) ~.
\end{eqnarray}
Remarkably, this leads to the compact expression 
\begin{eqnarray} 
\xi(i\sqrt{C_A}\gamma^2,m^2_q,m^2_q) &=& 4i 
\sqrt{C_A\gamma^4 + 4i\sqrt{C_A}\gamma^2m^2_q} \nonumber \\
&& \times \left[ \frac{\zeta(2)}{2} ~+~ \frac{1}{4} \ln^2 \left[ \frac
{\sqrt{C_A}\gamma^2 +\sqrt{C_A\gamma^4+4i\sqrt{C_A}\gamma^2m^2_q}}  
{\sqrt{C_A}\gamma^2 -\sqrt{C_A\gamma^4+4i\sqrt{C_A}\gamma^2m^2_q}}
\right] \right. \nonumber\\ 
&& \left. ~~~+~ \mbox{Li}_2\left[ \frac
{\sqrt{C_A\gamma^4+4i\sqrt{C_A}\gamma^2m^2_q} - \sqrt{C_A}\gamma^2}        
{\sqrt{C_A}\gamma^2 + \sqrt{C_A\gamma^4+4i\sqrt{C_A}\gamma^2m^2_q}} \right]
\right] 
\end{eqnarray} 
giving the integral to the finite part 
\begin{eqnarray}
I_2(m_q^2,m_q^2,i\sqrt{C_A}\gamma^2) &=& -~ \frac{1}{2 \epsilon^2} \left( 
i \sqrt{C_A} \gamma^2 ~+~ 2 m^2_q \right) \nonumber \\
&& -~ \frac{1}{\epsilon} \left(  \frac{1}{2}(3i \sqrt{C_A} \gamma^2 + 6m^2_q)
~-~2m^2_q\overline{\ln}(m^2_q) ~-~ i \sqrt{C_A} \gamma^2 
\overline{\ln}(i \sqrt{C_A} \gamma^2 ) \right) \nonumber \\
&& -~ m^2_q \overline{\ln}^2(m^2_q) ~-~ \frac{1}{2} i \sqrt{C_A} \gamma^2 
\overline{\ln}^2 (i \sqrt{C_A} \gamma^2) \nonumber \\ 
&& +~ 6m^2_q \overline{\ln}(m^2_q) ~+~ 3i \sqrt{C_A} \gamma^2 
\overline{\ln}(i \sqrt{C_A} \gamma^2 ) \nonumber \\ 
&& -~ 2i \sqrt{C_A \gamma^4 + 4i\sqrt{C_A} \gamma^2 m^2_q} \nonumber \\
&& ~~~ \times \left[ \frac{\zeta(2)}{2} ~+~ \frac{1}{4} \ln^2 \left[ 
\frac{ \sqrt{C_A} \gamma^2 + \sqrt{C_A \gamma^4 
+ 4i \sqrt{C_A} \gamma^2 m^2_q}}{ \sqrt{C_A} \gamma^2 
- \sqrt{C_A \gamma^4 + 4i \sqrt{C_A} \gamma^2 m^2_q}} \right] \right. 
\nonumber \\
&& \left. ~~~~~~~~+~ \mbox{Li}_2 \left[ \frac{ 
\sqrt{C_A \gamma^4 + 4i \sqrt{C_A} \gamma^2 m^2_q} - \sqrt{C_A} \gamma^2 } 
{ \sqrt{C_A \gamma^4 + 4i \sqrt{C_A} \gamma^2 m^2_q} + \sqrt{C_A} \gamma^2 } 
\right] \right] \nonumber \\  
&& -~ \frac{1}{2} \left( i \sqrt{C_A} \gamma^2 ~+~ 2m^2_q \right) 
\left(7 + \zeta(2) \right) 
\nonumber \\
&& -~ \frac{1}{2} \left( 2m^2_q - i \sqrt{C_A} \gamma^2 \right)
\overline{\ln}(m^2_q) \nonumber \\
&& ~-~ i\sqrt{C_A} \gamma^2 \overline{\ln}(m_q^2) \overline{\ln}(i\sqrt{C_A} 
\gamma^2) ~+~ O(\epsilon) ~.
\label{intp1} 
\end{eqnarray}  
We have checked that this expression correctly reduces to that for 
$I_2(0,0,i\sqrt{C_A}\gamma^2)$ in the limit $m_q^2$~$\rightarrow$~$0$. This is
not as straightforward as it seems due to the presence of the dilogarithm
function and terms involving $\overline{\ln}(m_q^2)$. Disregarding all terms 
proportional to $m^2_q$ and expanding 
$\sqrt{C_A\gamma^4+4i\sqrt{C_A}\gamma^2m^2_q}$ in powers of $m^2_q$, then 
$I_2(m_q^2,m_q^2,i\sqrt{C_A}\gamma^2)$ reduces to 
\begin{eqnarray}  
I_2(m_q^2,m_q^2,i\sqrt{C_A}\gamma^2) &=& -~ \frac{i \sqrt{C_A} \gamma^2}
{2 \epsilon^2} ~-~ \frac{i\sqrt{C_A} \gamma^2 }{\epsilon} \left(  \frac{3}{2} 
- \overline{\ln}(i \sqrt{C_A} \gamma^2 )\right) \nonumber \\
&& -~ \frac{1}{2} i \sqrt{C_A} \gamma^2 
\overline{\ln}^2(i \sqrt{C_A} \gamma^2) ~+~ 3i \sqrt{C_A} \gamma^2 
\overline{\ln}(i \sqrt{C_A} \gamma^2 ) \nonumber \\ 
&& -~ i \sqrt{C_A} \gamma^2 \left[ \zeta(2) ~+~ 
2\mbox{Li}_2 \left[\frac{im^2_q}{\sqrt{C_A} \gamma^2} \right] \right.
\nonumber \\
&& \left. ~~~~~~~~~~~~~~~+~ \frac{1}{2} \left[ \overline{\ln}^2 (i\sqrt{C_A} 
\gamma^2) - 2\overline{\ln}(i\sqrt{C_A} \gamma^2) \overline{\ln}(m^2_q)  
+ \overline{\ln}^2 (m^2_q) \right] 
\right] \nonumber \\
&& -~ \frac{i \sqrt{C_A} \gamma^2}{2} \left(7 + \zeta(2) \right) ~+~ 
\frac{i \sqrt{C_A} \gamma^2}{2} \overline{\ln}^2 (m^2_q) \nonumber \\
&& -~ i\sqrt{C_A} \gamma^2 \overline{\ln}(m_q^2) 
\overline{\ln}(i\sqrt{C_A} \gamma^2) ~+~ O(m_q^2;\epsilon) 
\end{eqnarray}  
where we note that the imaginary dilogarithm vanishes as 
$m^2_q$~$\rightarrow$~$0$ and the remaining logarithmic terms in $m^2_q$ 
cancel. By making the analytic continuation 
$m^2$~$\rightarrow$~$i\sqrt{C_A}\gamma^2$ in (\ref{int01}), we see that our 
integral $I_2(m_q^2,m_q^2,i\sqrt{C_A}\gamma^2)$ is entirely consistent with 
$I_2(0,0,i\sqrt{C_A}\gamma^2)$ in the limit of zero quark mass.
  
Next we turn to the complex conjugate integral and focus on
\begin{eqnarray} 
\xi(-i\sqrt{C_A}\gamma^2,m^2_q,m^2_q) &=& 8a \left[ 
2M(\phi_{m^2_q}) - M(-\phi_{-i\sqrt{C_A}\gamma^2}) \right] 
\end{eqnarray} 
where now the variables are 
\begin{equation}
a ~=~ \frac{i}{2} \sqrt{C_A\gamma^4 - 4i\sqrt{C_A}\gamma^2m^2_q} ~~~,~~~
c ~=~ 2m^2_q - i\sqrt{C_A}\gamma^2 
\end{equation}
leading to
\begin{eqnarray} 
e^{-\phi_{i\sqrt{C_A}\gamma^2}} &=& \frac
{\sqrt{C_A}\gamma^2 + \sqrt{C_A\gamma^4+4i\sqrt{C_A}\gamma^2m^2_q}} 
{\sqrt{C_A}\gamma^2 - \sqrt{C_A\gamma^4+4i\sqrt{C_A}\gamma^2m^2_q}}
\nonumber \\  
e^{-\phi_{m^2_q}} &=& \sqrt{\frac
{\sqrt{C_A}\gamma^2 - \sqrt{C_A\gamma^4+4i\sqrt{C_A}\gamma^2m^2_q}} 
{\sqrt{C_A}\gamma^2 + \sqrt{C_A\gamma^4+4i\sqrt{C_A}\gamma^2m^2_q}}} ~.
\end{eqnarray}  
Without reproducing analogous manipulations, we find
\begin{eqnarray} 
\xi(-i\sqrt{C_A}\gamma^2,m^2_q,m^2_q) &=& 
4i \sqrt{C_A\gamma^4 - 4i\sqrt{C_A}\gamma^2m^2_q} \nonumber \\
&& \times \left[ \frac{\zeta(2)}{2} ~+~ \frac{1}{4} \ln^2 \left[ \frac
{\sqrt{C_A}\gamma^2 -\sqrt{C_A\gamma^4-4i\sqrt{C_A}\gamma^2m^2_q}}  
{\sqrt{C_A}\gamma^2 +\sqrt{C_A\gamma^4-4i\sqrt{C_A}\gamma^2m^2_q}}
\right] \right. \nonumber\\ 
&& \left. ~~~~+~ \mbox{Li}_2\left[ \frac
{\sqrt{C_A\gamma^4-4i\sqrt{C_A}\gamma^2m^2_q} + \sqrt{C_A}\gamma^2}        
{\sqrt{C_A\gamma^4-4i\sqrt{C_A}\gamma^2m^2_q} - \sqrt{C_A}\gamma^2} \right] 
\right] ~. 
\end{eqnarray} 
Whilst this is similar to $\xi(i\sqrt{C_A}\gamma^2,m^2_q,m^2_q)$ there is a
potential singularity in the massless quark limit arising from the 
dilogarithm term. To circumvent this and to have a final expression for the
integral $I_2(-i\sqrt{C_A}\gamma^2,m^2_q,m^2_q)$ which is clearly the complex
conjugate of $I_2(i\sqrt{C_A}\gamma^2,m^2_q,m^2_q)$ we use the dilogarithm
identity, \cite{31}, 
\begin{eqnarray}
\mbox{Li}_2(-1/z) ~+~ \mbox{Li}_2(-z) &=& -~ \zeta(2) ~-~ \frac{1}{2}\ln^2(z) 
\end{eqnarray}    
with 
\begin{equation}
z ~=~ -~ \left[ \frac
{ \sqrt{C_A \gamma^4 - 4i \sqrt{C_A} \gamma^2 m^2_q} - \sqrt{C_A} \gamma^2 } 
{ \sqrt{C_A \gamma^4 - 4i \sqrt{C_A} \gamma^2 m^2_q} + \sqrt{C_A} \gamma^2 } 
\right] ~.
\end{equation}
Given this we end up with the final expression 
\begin{eqnarray}  
I_2(-i\sqrt{C_A}\gamma^2,m^2_q,m^2_q) &=& -~ \frac{1}{2 \epsilon^2} 
\left( 2 m^2_q ~-~i \sqrt{C_A} \gamma^2 \right) \nonumber \\
&& -~ \frac{1}{\epsilon} \left[  \frac{1}{2}( 6m^2_q - 3i \sqrt{C_A} \gamma^2 )
- 2m^2_q\overline{\ln}(m^2_q) + i \sqrt{C_A} \gamma^2 
\overline{\ln}(-i \sqrt{C_A} \gamma^2 ) \right]
\nonumber \\
&& -~ m^2_q (\overline{\ln}(m^2_q))^2 ~+~ \frac{1}{2} i \sqrt{C_A} \gamma^2 
(\overline{\ln}(-i \sqrt{C_A} \gamma^2))^2 
\nonumber \\ 
&& +~ 6m^2_q \overline{\ln}(m^2_q) ~-~ 3i \sqrt{C_A} \gamma^2 \overline{\ln}(-i \sqrt{C_A} 
\gamma^2 ) 
\nonumber \\ 
&& +~ 2i \sqrt{C_A \gamma^4 - 4i\sqrt{C_A} \gamma^2 m^2_q} 
 \nonumber \\
&& ~~~ \times \left[ \frac{\zeta(2)}{2} ~+~ 
\frac{1}{4} \ln^2 \left[ \frac
{ \sqrt{C_A} \gamma^2 - \sqrt{C_A \gamma^4 - 4i \sqrt{C_A} \gamma^2 m^2_q}}
{ \sqrt{C_A} \gamma^2 + \sqrt{C_A \gamma^4 - 4i \sqrt{C_A} \gamma^2 m^2_q}} 
\right] \right. \nonumber \\
&& \left. ~~~~~~~~+~ \mbox{Li}_2 \left[ \frac
{ \sqrt{C_A \gamma^4 - 4i \sqrt{C_A} \gamma^2 m^2_q} - \sqrt{C_A} \gamma^2 } 
{ \sqrt{C_A \gamma^4 - 4i \sqrt{C_A} \gamma^2 m^2_q} + \sqrt{C_A} \gamma^2 } 
\right] \right]  
\nonumber \\  
&& -~ \frac{1}{2} \left( 2m^2_q ~-~ i \sqrt{C_A} \gamma^2 \right) 
\left(7 + \zeta(2) \right) \nonumber \\
&& -~ \frac{1}{2} \left( 2m^2_q + i \sqrt{C_A} \gamma^2 \right)
\overline{\ln}^2(m^2_q) \nonumber \\
&& +~ i\sqrt{C_A} \gamma^2 \overline{\ln}(m_q^2) 
\overline{\ln}(-i\sqrt{C_A} \gamma^2) ~+~ O(\epsilon) ~.
\end{eqnarray}  
Comparing this with our expression, (\ref {intp1}), we see that the explicit
forms of $I_2(i\sqrt{C_A}\gamma^2,m^2_q,m^2_q)$ and
$I_2(-i\sqrt{C_A}\gamma^2,m^2_q,m^2_q)$ are indeed complex conjugates as 
expected from their original definitions. This is an important check on our
manipulations and use of dilogarithm identities and ensure that the correct 
massless quark limits will emerge which is important for checking our eventual 
gap equation.   

The remaining two master integrals, (\ref{mas2}), can be simply deduced from 
the above expressions by differentiating with respect to $\gamma^2$. As this
is elementary we merely note the explicit expression for the first of
(\ref{mas2}) is  
\begin{eqnarray} 
\bar{I}_2(i\sqrt{C_A}\gamma^2,m^2_q,m^2_q) &=& \frac{1}{2\epsilon^2} ~-~ 
\frac{1}{\epsilon} \left( \overline{\ln}(i\sqrt{C_A} \gamma^2) - \frac{1}{2} 
\right) \nonumber \\
&& +~ \frac{1}{2} \overline{\ln}^2 (i\sqrt{C_A} \gamma^2 ) ~-~ 
5 \overline{\ln}(i\sqrt{C_A} \gamma^2) + \frac{1}{2} + \frac{\zeta(2)}{2} ~+~
2i\pi \nonumber \\
&& -~ \frac{1}{2} \overline{\ln}^2 (m^2_q)  ~+~ \overline{\ln}(m^2_q) 
\overline{\ln}(i\sqrt{C_A} \gamma^2) ~-~ 4\ln(2) \nonumber \\
&& +~ \left[ \frac{\left( 2\sqrt{C_A} \gamma^2 + 4im^2_q\right)  
\sqrt{C_A \gamma^4 - 4i\sqrt{C_A} \gamma^2 m^2_q}}  
{\sqrt{C^2_A \gamma^8 + 16\sqrt{C_A}^2 \gamma^4 m^4_q}} \right] \nonumber \\
&& ~~~ \times  \left[ \frac{\zeta(2)}{2} ~+~ 
\frac{1}{4} \ln^2 \left[ \frac{ \sqrt{C_A} \gamma^2 
+ \sqrt{C_A \gamma^4 + 4i \sqrt{C_A} \gamma^2 m^2_q}}{ \sqrt{C_A} \gamma^2 
- \sqrt{C_A \gamma^4 + 4i \sqrt{C_A} \gamma^2 m^2_q}} \right] 
\right. \nonumber \\
&& \left. ~~~~~~~~+~ \mbox{Li}_2 \left[ \frac{ \sqrt{C_A \gamma^4 
+ 4i \sqrt{C_A} \gamma^2 m^2_q} - \sqrt{C_A} \gamma^2 }
{ \sqrt{C_A \gamma^4 + 4i \sqrt{C_A} \gamma^2 m^2_q} + \sqrt{C_A} \gamma^2 } 
\right] \right] \nonumber \\  
&& ~~~~~~~~+~ 4\overline{\ln} \left[ \sqrt{C_A} \gamma^2 
+ \sqrt{C_A \gamma^4 + 4i \sqrt{C_A} \gamma^2 m^2_q} \right] \,+\, O(\epsilon)
\end{eqnarray}
where we have used 
\begin{eqnarray}
\frac{d~}{dz} \mbox{Li}_2(z) ~=~ -~ \frac{\ln(1-z)}{z} ~.
\end{eqnarray}
Again we have checked that the correct massless quark limit emerges with the
direct evaluation of the equivalent integral. 

Whilst we have now determined all the master integrals to the finite part in
the $\epsilon$ expansion, the explicit expressions are not in a fully useful
format. Given that the ultimate gap equation is a real function we need to 
write the expressions as a real and imaginary part. This is not a simple
exercise due to the presence of the dilogarithm of a complex argument.
However, the theory behind such functions is known, \cite{31}, and we summarize
what we require for the current calculation. Writing the complex variable $z$
in polar form we have the real and imaginary parts, \cite{31}, 
\begin{equation}
\mbox{Li}_2(re^{i\theta}) ~=~ \mbox{Li}_2(r,\theta) ~+~ i \left[ \omega \ln(r)
+ \half \mbox{Cl}_2(2\omega) + \half \mbox{Cl}_2 (2\theta) - \half 
\mbox{Cl}_2(2\omega + 2\theta) \right]
\end{equation} 
where 
\begin{equation}
\mbox{Li}_2(r,\theta) ~=~ -~ \frac{1}{2}\int_0^r 
\frac{\ln(1-2x\cos\theta +x^2)}{x} dx 
\end{equation} 
and $\mbox{Cl}_2(\theta)$ is the Clausen function defined by
\begin{equation}
\mbox{Cl}_2(\theta) ~=~ -~ \int_0^\theta \ln 
\left[2\sin\left(\frac{\phi}{2}\right)\right] d\phi ~. 
\end{equation}   
The intermediate angle $\omega$ is related to the polar variables $r$ and 
$\theta$ of $z$ by
\begin{equation}
\omega ~=~ \tan^{-1} \left( \frac{r\sin\theta}{1-r\cos\theta} \right) ~.
\end{equation}
Given these general definitions then to proceed with our simplification to real
and imaginary parts, we need to write the arguments of the dilogarithms in 
polar forms. To assist this we recall the elementary lemma for a complex 
variable $z$~$=$~$a$~$+$~$ib$, where $a$ and $b$ are real, 
\begin{equation}
\sqrt{a \pm ib} ~=~ \frac{1}{\sqrt{2}}\sqrt{\sqrt{a^2+b^2} + a} ~\pm~
\frac{i}{\sqrt{2}}\sqrt{\sqrt{a^2+b^2} - a} ~.
\end{equation}
So, for example, 
\begin{eqnarray}
\sqrt{C_A \gamma^4 + 4i\sqrt{C_A} \gamma^2 m^2_q} &=& \frac{1}{\sqrt{2}} 
\sqrt{\sqrt{C_A^2 \gamma^8+16C_A\gamma^4m^2_q} + C_A\gamma^4} \nonumber \\ 
&& +~ \frac{i}{\sqrt{2}}\sqrt{\sqrt{C_A^2 \gamma^8+16C_A\gamma^4m^2_q} 
- C_A\gamma^4} ~.
\end{eqnarray}
For the dilogarithms if we set 
\begin{equation} 
re^{i\theta} ~\equiv~ \frac{ \sqrt{C_A \gamma^4 + 4i \sqrt{C_A} \gamma^2 m^2_q}
- \sqrt{C_A} \gamma^2 }{ \sqrt{C_A \gamma^4 + 4i \sqrt{C_A} \gamma^2 m^2_q} 
+ \sqrt{C_A} \gamma^2 } 
\end{equation} 
then
\begin{equation}
re^{i\theta} ~=~ \frac{\sqrt{C_A^2 \gamma^8+16C_A\gamma^4m^2_q} - C_A\gamma^4 
+ i\sqrt{2}\sqrt{C_A} \gamma^2 
\sqrt{\sqrt{C_A^2 \gamma^8+16C_A\gamma^4m^2_q} - C_A\gamma^4}} 
{\sqrt{C_A^2 \gamma^8+16C_A\gamma^4m^2_q} + C_A\gamma^4 + \sqrt{2}\sqrt{C_A} 
\gamma^2 \sqrt{\sqrt{C_A^2 \gamma^8+16C_A\gamma^4m^2_q} + C_A\gamma^4}} ~. 
\end{equation} 
giving 
\begin{eqnarray}
r &=& \frac{4\sqrt{C_A}\gamma^2m^2_q}
{\sqrt{\sqrt{C_A^2 \gamma^8+16C_A\gamma^4m^2_q} + C_A\gamma^4} 
\left( \sqrt{2} \sqrt{C_A} \gamma^2 + 
\sqrt{\sqrt{C_A^2 \gamma^8+16C_A\gamma^4m^2_q} + C_A\gamma^4} \right)} 
\nonumber \\ 
\tan \theta &=& \frac{ \sqrt{2} \sqrt{C_A} \gamma^2}  
{\sqrt{\sqrt{C_A^2 \gamma^8+16C_A\gamma^4m^2_q} - C_A\gamma^4}} ~.
\end{eqnarray}
In what follows we will always regard $r$ and $\theta$ as taking these values
with the associated corresponding value of $\omega$. Although the dilogarithm 
is the most involved of the terms which appear in the finite parts, similar 
manipulation is required for several of the logarithm terms. Collecting all the
pieces together we find the following expression written as real and imaginary 
parts,
\begin{eqnarray}
I_2(i\sqrt{C_A}\gamma^2,m^2_q,m^2_q) &=& -~ \frac{1}{2 \epsilon^2} 
\left( i \sqrt{C_A} \gamma^2 ~+~ 2 m^2_q \right) \nonumber \\
&& -~ \frac{1}{\epsilon} \left( \frac{1}{2}(3i \sqrt{C_A} \gamma^2 + 6m^2_q)
- 2m^2_q\overline{\ln}(m^2_q) - i \sqrt{C_A} \gamma^2 \overline{\ln}
(i \sqrt{C_A} \gamma^2 ) \right) \nonumber \\
&& -~ m^2_q \overline{\ln}^2 (m^2_q) ~-~ \frac{1}{2} i \sqrt{C_A} \gamma^2 
\overline{\ln}^2 (i \sqrt{C_A} \gamma^2) \nonumber \\ 
&& +~ 6m^2_q \overline{\ln}(m^2_q) ~+~ 3i \sqrt{C_A} \gamma^2 
\overline{\ln}(i \sqrt{C_A} \gamma^2 ) \nonumber \\ 
&& -~ \sqrt{2} i \sqrt{ \sqrt{C^2_A \gamma^8 + 16C_A \gamma^4 m^2_q } 
+ C_A \gamma^4 } \nonumber \\
&& ~~~ \times \left[ \frac{1}{4} \left[ \frac{1}{2}\overline{\ln} \left( 
\sqrt{C^2_A \gamma^8 + 16C_A \gamma^4 m^4_q } + C_A \gamma^4 \right) 
\right. \right. \nonumber \\
&& \left. \left. ~~~~~~~~~~~~ \times \overline{\ln} \left( \sqrt{2} \sqrt{C_A} 
\gamma^2 + \sqrt{ \sqrt{C^2_A \gamma^8 + 16 C_A \gamma^4 m^4_q } 
+ C_A \gamma^4} \right) \right. \right. \nonumber \\ 
&& \left. \left. ~~~~~~~~~~~~+~ 2i\tan^{-1} \!\! \left[ 
\frac{\sqrt{ \sqrt{C^2_A \gamma^8 + 16 C_A \gamma^4 m^4_q } - C_A \gamma^4}}
{\sqrt{2} \sqrt{C_A} \gamma^2 + \!\! \sqrt{ \! \sqrt{C^2_A \gamma^8 
+ 16 C_A \gamma^4 m^4_q } + C_A \gamma^4}} \right] \right. \right. \nonumber \\ 
&& \left. \left. ~~~~~~~~~~~~-~ 2\ln(2) + \frac{1}{2} i\pi 
- \overline{\ln}(\sqrt{C_A} \gamma^2) - \overline{\ln}(m^2_q) 
+ \frac{\zeta(2)}{2} \right]^2 \right. 
\nonumber \\
&& \left. ~~~~~~~+~ \mbox{Li}_2(r,\theta) \right. \nonumber \\
&& \left. ~~~~~~~+~ i \omega \left[ 2\ln(2) + \frac{1}{2} 
\overline{\ln} (C_A \gamma^4) + \overline{\ln}(m^2_q) \right. \right. 
\nonumber \\ 
&& \left. \left. ~~~~~~~~~~~~~~~~-~ \frac{1}{2} \overline{\ln}\left( \sqrt{ 
C^2_A \gamma^8 + 16C_A \gamma^4 m^4_q } + C_A \gamma^4 \right) \right. \right.
\nonumber \\ 
&& \left. \left. ~~~~~~~~~~~~~~~~-~ \overline{\ln}\left( \sqrt{2} \sqrt{C_A} 
\gamma^2 + \sqrt{ \sqrt{ C^2_A \gamma^8 + 16C_A \gamma^4 m^4_q } 
+ C_A \gamma^4} \right) \right] \right. \nonumber \\
&& \left. ~~~~~~~+~ \frac{i}{2} \mbox{Cl}_2 (2\omega)
+ \frac{i}{2} \mbox{Cl}_2 (2\theta)
- \frac{i}{2} \mbox{Cl}_2 (2\omega + 2\theta) \right]   
\nonumber \\ 
&& +~ \sqrt{2} \sqrt{ \sqrt{C^2_A \gamma^8 + 16C_A \gamma^4 m^2_q } 
- C_A \gamma^4 } \nonumber \\
&& ~~~ \times \left[ \frac{1}{4} \left[ \frac{1}{2}\overline{\ln} \left( 
\sqrt{C^2_A \gamma^8 + 16C_A \gamma^4 m^4_q } + C_A \gamma^4 \right) 
\right. \right. \nonumber \\
&& \left. \left. ~~~~~~~~~~~~ \times \overline{\ln} \left( \sqrt{2} \sqrt{C_A}
\gamma^2 + \sqrt{ \sqrt{C^2_A \gamma^8 + 16 C_A \gamma^4 m^4_q } 
+ C_A \gamma^4} \right) \right. \right. \nonumber \\ 
&& \left. \left. ~~~~~~~~~~~~+~ 2i\tan^{-1} \!\! \left[ 
\frac{\sqrt{ \sqrt{C^2_A \gamma^8 + 16 C_A \gamma^4 m^4_q } - C_A \gamma^4}}
{\sqrt{2} \sqrt{C_A} \gamma^2 + \!\! \sqrt{ \!\! \sqrt{C^2_A \gamma^8 
+ 16 C_A \gamma^4 m^4_q } + C_A \gamma^4}} \right]   
\right. \right. \nonumber \\
&& \left. \left. ~~~~~~~~~~~~-~ 2\ln(2) + \frac{1}{2} i\pi 
- \overline{\ln}(\sqrt{C_A} \gamma^2) - \overline{\ln}(m^2_q) 
+ \frac{\zeta(2)}{2} \right]^2 \right. \nonumber \\
&& \left. ~~~~~~~+~ \mbox{Li}_2(r,\theta) \right. \nonumber \\
&& \left. ~~~~~~~+~ i \omega \left[ 2\ln(2) + \frac{1}{2} 
\overline{\ln} (C_A \gamma^4) + \overline{\ln}(m^2_q) \right. \right. 
\nonumber \\ 
&& \left. \left. ~~~~~~~~~~~~~~~~-~ \frac{1}{2} \overline{\ln}\left( \sqrt{ 
C^2_A \gamma^8 + 16C_A \gamma^4 m^4_q } + C_A \gamma^4 \right) \right. \right. 
\nonumber \\ 
&& \left. \left. ~~~~~~~~~~~~~~~~-~ \overline{\ln}\left( \sqrt{2} \sqrt{C_A} 
\gamma^2 + \sqrt{ \sqrt{ C^2_A \gamma^8 + 16C_A \gamma^4 m^4_q } 
+ C_A \gamma^4} \right) \right] \right. \nonumber \\
&& \left. ~~~~~~~+~ \frac{i}{2} \mbox{Cl}_2 (2\omega)
+ \frac{i}{2} \mbox{Cl}_2 (2\theta)
- \frac{i}{2} \mbox{Cl}_2 (2\omega + 2\theta) \right]   
\nonumber \\  
&& -~ \frac{1}{2} \left( i \sqrt{C_A} \gamma^2 ~+~ 2m^2_q \right) 
\left(7 + \zeta(2) \right) ~-~ \frac{1}{2} \left( 2m^2_q - i \sqrt{C_A} 
\gamma^2 \right) \overline{\ln}^2 (m^2_q) \nonumber \\
&& -~ i\sqrt{C_A} \gamma^2 \overline{\ln}(m^2_q) \overline{\ln}(i\sqrt{C_A} 
\gamma^2) ~+~ O(\epsilon) ~.
\end{eqnarray}  
Whilst this is not truly of the form $a$~$+$~$ib$ since not all terms have been
fully multiplied out and there are logarithms with purely imaginary arguments,
we prefer to leave it in this more compact form since, for instance, it is
elementary to implement
\begin{equation}
\overline{\ln} (i\sqrt{C_A}\gamma^2) ~=~ 
\overline{\ln} (\sqrt{C_A}\gamma^2) ~+~ \frac{i \pi}{2} 
\end{equation} 
within our {\sc Form} routines. This also takes care of the other elementary 
complex algebra automatically. 

\sect{Two loop gap equation.}

Equipped with the basic master integrals we are now in a position to assemble
the two loop Gribov gap equation in the $\MSbar$ scheme with massive quarks.
This requires the evaluation of the seventeen contributing Feynman diagrams
which without the power of {\sc Form} would have been virtually impossible. 
Having already discussed the key aspects of the computation, we ultimately find 
\begin{eqnarray}
1 &=& aC_A \left[ \frac{5}{8}~-~ \frac{3}{8} \ln \left( 
\frac{C_A \gamma^4}{\mu^4} \right) \right] \nonumber \\
&& +~ a^2 \left( \frac{\sqrt{C_A} T_F \Nf m^2_q}{\gamma^2}\right) 
\left[4\omega+\frac{\pi}{2} \right]  
\nonumber \\
&& +~ a^2\left[ C^2_A \left[ \frac{2017}{768} - \frac{11097}{2048}s_2 
+ \frac{95}{256}\zeta(2) - \frac{65}{48} \overline{\ln}(C_A \gamma^4) 
\right. \right. \nonumber \\   
&& \left. \left. ~~~~~~+~ \frac{35}{128}\left(\overline{\ln}(C_A \gamma^4) 
\right)^2 + \frac{1137}{2560} \sqrt{5} \zeta(2) - \frac{205 \pi^2}{512} \right] 
\right. \nonumber \\
&& \left. ~~~~~~+~ C_A T_F \Nf \left[ 2 \ln(2) - \frac{25}{24} 
+\frac{1}{2} \overline{\ln}^2 (m^2_q) 
-\frac{1}{2} \overline{\ln}(m^2_q) \overline{\ln}(C_A\gamma^4) 
\right. \right. \nonumber \\
&& \left. \left. ~~~~~~~~~~~~~~~~~~~~~~~+~ \frac{19}{12} 
\overline{\ln}(C_A \gamma^4) - \overline{\ln} 
\left[\sqrt{\sqrt{C_A^2\gamma^8+16C_A\gamma^4m^4_q}+C_A\gamma^4}\right]
\right. \right. \nonumber \\ 
&& \left. \left. ~~~~~~-~ \overline{\ln} \left[\sqrt{2}\sqrt{C_A}\gamma^2 + 
\sqrt{\sqrt{C_A^2\gamma^8+16C_A\gamma^4m^4_q}+C_A\gamma^4}\right]  
+ \frac{\pi^2}{8} \right] \right] \nonumber \\
&& +~ a^2 \sqrt{\sqrt{C_A^2 \gamma^8 + 16C_A\gamma^4 m^2_q} + C_A\gamma^4} 
\left( \frac{\sqrt{C_A} T_F \Nf}{\sqrt{2} \gamma^2}\right) \nonumber \\
&& ~~~ \times \left[ \,-~ \frac{\zeta(2)}{4} - \frac{1}{2} \ln^2(2) 
- \frac{1}{2} \ln(2) \overline{\ln}(m^2_q) - \frac{1}{4} \ln(2) 
\ln(C_A \gamma^4) \right. \nonumber \\ 
&& \left. ~~~~~~~+~ \frac{1}{2} \ln(2) \overline{\ln} \left[ \sqrt{2} 
\sqrt{C_A} \gamma^2 + \sqrt{\sqrt{C_A^2 \gamma^8 + 16C_A \gamma^2 m^4_q} 
+ C_A\gamma^4} \right] \right. \nonumber \\ 
&& \left. ~~~~~~~+~ \frac{1}{2} \ln(2) \overline{\ln} \left[ 
\sqrt{\sqrt{C_A^2 \gamma^8 + 16C_A \gamma^2 m^4_q} + C_A\gamma^4} \right] 
- \frac{1}{8} \overline{\ln}^2 (m^2_q) - \frac{1}{8} \overline{\ln}(m^2_q) 
\overline{\ln}(C_A \gamma^4) \right. \nonumber \\ 
&& \left. ~~~~~~~+~ \frac{1}{4} \overline{\ln}(m^2_q) \overline{\ln} \left[ 
\sqrt{2} \sqrt{C_A} \gamma^2 + \sqrt{\sqrt{C_A^2 \gamma^8 
+ 16C_A\gamma^4m^4_q} + C_A\gamma^4} \right] 
- \frac{1}{32} \overline{\ln}^2 (C_A \gamma^4) \right.
\nonumber \\
&& \left. ~~~~~~~+~ \frac{1}{8} \overline{\ln}(C_A\gamma^4) \overline{\ln} 
\left[ \sqrt{2}\sqrt{C_A} \gamma^2 + \sqrt{\sqrt{C_A^2 \gamma^8 
+ 16C_A\gamma^4m^4_q} + C_A\gamma^4 } \right] \right. 
\nonumber \\
&& \left. ~~~~~~~-~ \frac{1}{8} \overline{\ln}^2 \left[ \sqrt{2}\sqrt{C_A} 
\gamma^2 + \sqrt{\sqrt{C_A^2 \gamma^8 + 16C_A\gamma^4m^4_q} + C_A\gamma^4 } 
\right] \right. \nonumber \\  
&& \left. ~~~~~~~+~ \frac{1}{4}  \overline{\ln}(m^2_q) \overline{\ln} \left[ 
\sqrt{\sqrt{C_A^2 \gamma^8 + 16C_A\gamma^4m^4_q} + 
C_A\gamma^4 } \right]  
\right. \nonumber \\  
&& \left. ~~~~~~~+~ \frac{1}{8}  \overline{\ln}(\sqrt{C_A}\gamma^2) 
\overline{\ln} \left[ \sqrt{\sqrt{C_A^2 \gamma^8 + 16C_A\gamma^4m^4_q} + 
C_A\gamma^4 } \right] \right. \nonumber \\
&& \left. ~~~~~~~-~ \frac{1}{4} \overline{\ln} 
\left[\sqrt{\sqrt{C_A^2\gamma^8+16C_A\gamma^4m^4_q}+C_A\gamma^4}\right]
\right. \nonumber \\
&& \left. ~~~~~~~~~~~ \times \overline{\ln} \left[\sqrt{2}\sqrt{C_A}\gamma^2 + 
\sqrt{\sqrt{C_A^2\gamma^8+16C_A\gamma^4m^4_q}+C_A\gamma^4}\right]
\right. \nonumber \\
&& \left. ~~~~~~~~-~ \frac{1}{8}   
\overline{\ln}^2 
\left[\sqrt{\sqrt{C_A^2\gamma^8+16C_A\gamma^4m^4_q}+C_A\gamma^4}\right]
+\frac{\pi}{4}\omega + \frac{\pi^2}{32} \right] \nonumber \\  
&& +~ a^2 \left[ \frac{\sqrt{\sqrt{C_A^2 \gamma^8 + 16C_A\gamma^4 m^2_q} 
+ C_A\gamma^4}}{\sqrt{C_A^2\gamma^8+16C_A\gamma^4m^4_q}} \right] 
\left( \frac{\sqrt{C_A} T_F \Nf m^4_q}{\sqrt{2} \gamma^2}\right) \nonumber \\
&& ~~~ \times \left[ \,-~ \zeta(2) - 2 \ln^2(2) - 2\ln(2)\overline{\ln}(m^2_q) 
- \ln(2)\overline{\ln}(C_A\gamma^4) \right. \nonumber \\
&& \left. ~~~~~~~+~ 2 \ln(2) \overline{\ln} \left[\sqrt{2}\sqrt{C_A}\gamma^2 + 
\sqrt{\sqrt{C_A^2\gamma^8+16C_A\gamma^4m^4_q}+C_A\gamma^4}\right]  
\right. \nonumber \\
&& \left. ~~~~~~~+~ 2 \ln(2) \overline{\ln} 
\left[\sqrt{\sqrt{C_A^2\gamma^8+16C_A\gamma^4m^4_q}+C_A\gamma^4}\right]
- \frac{1}{2} \overline{\ln}^2 (m^2_q)
- \frac{1}{2} \overline{\ln}(m^2_q) \overline{\ln}(C_A\gamma^4) \right. 
\nonumber \\
&& \left. ~~~~~~~+~ \overline{\ln}(m^2_q) 
\overline{\ln} \left[\sqrt{2}\sqrt{C_A}\gamma^2 + 
\sqrt{\sqrt{C_A^2\gamma^8+16C_A\gamma^4m^4_q}+C_A\gamma^4}\right]  
\right. \nonumber\\ 
&& \left. ~~~~~~~-~ \frac{1}{8} \overline{\ln}^2 (C_A\gamma^4) 
+ \frac{1}{2} \overline{\ln}(C_A\gamma^4) \overline{\ln} 
\left[\sqrt{2}\sqrt{C_A}\gamma^2 
+ \sqrt{\sqrt{C_A^2\gamma^8+16C_A\gamma^4m^4_q}+C_A\gamma^4}\right] \right. 
\nonumber \\
&& \left. ~~~~~~~-~ \frac{1}{2} \overline{\ln}^2
\left[\sqrt{2}\sqrt{C_A}\gamma^2 
+ \sqrt{\sqrt{C_A^2\gamma^8+16C_A\gamma^4m^4_q}+C_A\gamma^4}\right]  
\right. \nonumber \\
&& \left. ~~~~~~~+~ \overline{\ln}(m^2_q) 
\overline{\ln} \left[
\sqrt{\sqrt{C_A^2\gamma^8+16C_A\gamma^4m^4_q}+C_A\gamma^4}\right]
\right. \nonumber \\
&& \left. ~~~~~~~+~ \frac{1}{2}\overline{\ln}(C_A\gamma^4) 
\overline{\ln} \left[
\sqrt{\sqrt{C_A^2\gamma^8+16C_A\gamma^4m^4_q}+C_A\gamma^4}\right]
\right. \nonumber \\
&& \left. ~~~~~~~-~ \overline{\ln} \left[
\sqrt{\sqrt{C_A^2\gamma^8+16C_A\gamma^4m^4_q}+C_A\gamma^4}\right]
\right. \nonumber \\
&& \left. ~~~~~~~~~~~ \times \overline{\ln} \left[\sqrt{2}\sqrt{C_A}\gamma^2 
+ \sqrt{\sqrt{C_A^2\gamma^8+16C_A\gamma^4m^4_q}+C_A\gamma^4}\right]  
\right. \nonumber \\
&& \left. ~~~~~~~-~ \frac{1}{2} \overline{\ln}^2 
\left[\sqrt{\sqrt{C_A^2\gamma^8+16C_A\gamma^4m^4_q}+C_A\gamma^4}\right]
+ 2\omega^2 - 2\mbox{Li}_2(r,\theta) +\pi\omega 
+ \frac{\pi^2}{8}\right] \nonumber \\ 
&& +~ a^2 \left[ \frac{\sqrt{\sqrt{C_A^2 \gamma^8 + 16C_A\gamma^4 m^2_q} 
+ C_A\gamma^4}}{\sqrt{C_A^2\gamma^8+16C_A\gamma^4m^4_q}} \right] 
\left( \frac{(C_A)^{3/2} T_F \Nf \gamma^2}{\sqrt{2}}\right) \nonumber \\
&& ~~~ \times \left[ \,-~ \frac{\zeta(2)}{4} -\frac{1}{2} \ln^2(2) 
- \frac{1}{2}\ln(2)\overline{\ln}(m^2_q) 
- \frac{1}{4}\ln(2)\overline{\ln}(C_A\gamma^4) \right. \nonumber \\
&& \left. ~~~~~~~+~ \frac{1}{2} \ln(2)  
\overline{\ln} \left[\sqrt{2}\sqrt{C_A}\gamma^2 + 
\sqrt{\sqrt{C_A^2\gamma^8+16C_A\gamma^4m^4_q}+C_A\gamma^4}\right]  
\right. \nonumber \\
&& \left. ~~~~~~~+~ \frac{1}{2} \ln(2) \overline{\ln} 
\left[\sqrt{\sqrt{C_A^2\gamma^8+16C_A\gamma^4m^4_q}+C_A\gamma^4}\right]
- \frac{1}{8} \overline{\ln}^2 (m^2_q) 
- \frac{1}{8} \overline{\ln}(m^2_q) \overline{\ln}(C_A\gamma^4) \right. 
\nonumber \\
&& \left. ~~~~~~~+~ \frac{1}{4} \overline{\ln}(m^2_q) 
\overline{\ln} \left[\sqrt{2}\sqrt{C_A}\gamma^2 + 
\sqrt{\sqrt{C_A^2\gamma^8+16C_A\gamma^4m^4_q}+C_A\gamma^4}\right] 
- \frac{1}{32} \overline{\ln}^2 (C_A\gamma^4) \right. 
\nonumber \\
&& \left. ~~~~~~~+~ \frac{1}{8} \overline{\ln}(C_A\gamma^4) 
\overline{\ln} \left[\sqrt{2}\sqrt{C_A}\gamma^2 + 
\sqrt{\sqrt{C_A^2\gamma^8+16C_A\gamma^4m^4_q}+C_A\gamma^4}\right]  
\right. \nonumber \\
&& \left. ~~~~~~~-~ \frac{1}{8} \overline{\ln}^2 
\left[\sqrt{2}\sqrt{C_A}\gamma^2 
+ \sqrt{\sqrt{C_A^2\gamma^8+16C_A\gamma^4m^4_q}+C_A\gamma^4}\right]  
\right. \nonumber \\
&& \left. ~~~~~~~+~ \frac{1}{4} \overline{\ln}(m^2_q) 
\overline{\ln} \left[
\sqrt{\sqrt{C_A^2\gamma^8+16C_A\gamma^4m^4_q}+C_A\gamma^4}\right]
\right. \nonumber \\
&& \left. ~~~~~~~+~ \frac{1}{8} \overline{\ln}(C_A\gamma^4) \overline{\ln} 
\left[\sqrt{\sqrt{C_A^2\gamma^8+16C_A\gamma^4m^4_q}+C_A\gamma^4}\right]
\right. \nonumber \\
&& \left. ~~~~~~~-~ \frac{1}{4} \overline{\ln} 
\left[\sqrt{\sqrt{C_A^2\gamma^8+16C_A\gamma^4m^4_q}+C_A\gamma^4}\right]
\right. \nonumber\\
&& \left. ~~~~~~~~~~~ \times \overline{\ln} \left[\sqrt{2}\sqrt{C_A}\gamma^2 
+ \sqrt{\sqrt{C_A^2\gamma^8+16C_A\gamma^4m^4_q}+C_A\gamma^4}\right]  
\right. \nonumber \\ 
&& \left. ~~~~~~~-~ \frac{1}{8} \overline{\ln}^2 
\left[\sqrt{\sqrt{C_A^2\gamma^8+16C_A\gamma^4m^4_q}+C_A\gamma^4}\right]
+ \frac{\omega^2}{2} -\frac{1}{2}\mbox{Li}_2(r,\theta) 
+ \frac{\pi\omega}{4} + \frac{\pi^2}{32}  \right] \nonumber \\  
&& +~ a^2 \sqrt{\sqrt{C_A^2 \gamma^8 + 16C_A\gamma^4 m^2_q} - C_A\gamma^4} 
\left( \frac{\sqrt{C_A} T_F \Nf}{\sqrt{2}\gamma^2}\right) \nonumber \\
&& ~~~ \times \left[ \,-~ \frac{\pi}{4}\ln(2) 
- \frac{\pi}{8}\overline{\ln}(m^2_q) 
- \frac{\pi}{16}\overline{\ln}(C_A\gamma^4) + \frac{\pi}{8} \overline{\ln} 
\left[\sqrt{\sqrt{C_A^2\gamma^8+16C_A\gamma^4m^4_q}+C_A\gamma^4}
\right] \right. \nonumber \\
&& \left. ~~~~~~~+~ \frac{\pi}{8}
\overline{\ln} \left[\sqrt{2}\sqrt{C_A}\gamma^2 + 
\sqrt{\sqrt{C_A^2\gamma^8+16C_A\gamma^4m^4_q}+C_A\gamma^4}\right]  
\right. \nonumber \\ 
&& \left. ~~~~~~~+~ \frac{1}{4}\mbox{Cl}_2(2\theta) 
- \frac{1}{4}\mbox{Cl}_2(2\theta+2\omega) + \frac{1}{4}\mbox{Cl}_2(2\omega)
\right. \nonumber \\
&& \left. ~~~~~~~+~ \left[\frac{m^4_q}{\sqrt{C_A^2\gamma^8+16C_A\gamma^4m^4_q}}
\right] 
\right. \nonumber \\  
&& \left. ~~~~~~~~~~~ \times \left[ \pi\ln(2) 
+ \frac{\pi}{2}\overline{\ln}(m^2_q) +\frac{\pi}{4}\overline{\ln}(C_A\gamma^4) 
- \frac{1}{2} \overline{\ln} \left[
\sqrt{\sqrt{C_A^2\gamma^8+16C_A\gamma^4m^4_q}+C_A\gamma^4}\right]
\right. \right. \nonumber \\
&& \left. \left. ~~~~~~~~~~~~~~~-~ \frac{1}{2} \overline{\ln} 
\left[\sqrt{2}\sqrt{C_A}\gamma^2 
+ \sqrt{\sqrt{C_A^2\gamma^8+16C_A\gamma^4m^4_q}+C_A\gamma^4}\right]  
\right. \right. \nonumber \\
&& \left. \left. ~~~~~~~~~~~~~~~-~ \mbox{Cl}_2(2\theta) 
+ \mbox{Cl}_2(2\theta+ 2\omega) - \mbox{Cl}_2(2\omega) \right] \right] 
\nonumber \\  
&& +~ a^2 \left[\frac{\sqrt{\sqrt{C_A^2 \gamma^8 + 16C_A\gamma^4 m^4_q} 
- C_A\gamma^4}}{\sqrt{C_A^2\gamma^8 + 16C_A\gamma^4m^4_q}}\right]  
\left( \frac{(C_A)^{3/2}\gamma^2 T_F \Nf}{\sqrt{2}}\right) \nonumber \\
&& ~~~ \times \left[ \frac{1}{4} \mbox{Cl}_2(2\theta+2\omega) 
- \frac{1}{4} \mbox{Cl}_2(2\theta) - \frac{1}{4} \mbox{Cl}_2(2\omega) 
+ \frac{\pi}{4}\ln(2) + \frac{\pi}{8} \overline{\ln}(m^2_q) \right. 
\nonumber \\ 
&& \left. ~~~~~~~+~ \frac{\pi}{16} \overline{\ln}(C_A\gamma^4) 
- \frac{\pi}{8} \overline{\ln} 
\left[\sqrt{\sqrt{C_A^2\gamma^8+16C_A\gamma^4m^4_q}+C_A\gamma^4}\right]
\right. \nonumber\\ 
&& \left. ~~~~~~~-~ \frac{1}{8} \overline{\ln} 
\left[\sqrt{2}\sqrt{C_A}\gamma^2 
+ \sqrt{\sqrt{C_A^2\gamma^8+16C_A\gamma^4m^4_q}+C_A\gamma^4}\right] \right] ~+~
O(a^3) ~. 
\label{gap2}
\end{eqnarray}
This is a real expression and the main result of our article. There are several
checks. Whilst we have been careful in checking that the four two loop scalar 
master integrals reduce to the correct expressions in the massless quark limit,
the overall final gap equation must also satisfy the same test. We note that 
(\ref{gap2}) does do this and for completeness note that one obtains 
\begin{eqnarray} 
1 &=& C_A \left[ \frac{5}{8} - \frac{3}{8} \ln \left( 
\frac{C_A\gamma^4}{\mu^4} \right) \right] a \nonumber \\ 
&& +~ \left[ C_A^2 \left( \frac{2017}{768} - \frac{11097}{2048} s_2
+ \frac{95}{256} \zeta(2)
- \frac{65}{48} \ln \left( \frac{C_A\gamma^4}{\mu^4} \right)
+ \frac{35}{128} \left( \ln \left( \frac{C_A\gamma^4}{\mu^4} \right)
\right)^2 \right. \right. \nonumber \\
&& \left. \left. ~~~~~~~~~~~~+~ \frac{1137}{2560} \sqrt{5} \zeta(2) 
- \frac{205\pi^2}{512} \right) \right. \nonumber \\
&& \left. ~~~~~+~ C_A T_F \Nf \left( -~ \frac{25}{24} - \zeta(2)
+ \frac{7}{12} \ln \left( \frac{C_A\gamma^4}{\mu^4} \right)
- \frac{1}{8} \left( \ln \left( \frac{C_A\gamma^4}{\mu^4} \right) \right)^2 
+ \frac{\pi^2}{8} \right) \right] a^2 \nonumber \\
&& +~ O(a^3)  
\label{gap20}
\end{eqnarray} 
where $s_2$ $=$ $(2\sqrt{3}/9) \mbox{Cl}_2(2\pi/3)$ which was originally
recorded in \cite{15}. However, there is another check on (\ref{gap2}) which is
to examine the Faddeev-Popov ghost $2$-point function in the zero momentum 
limit. As was noted in \cite{1} there ought to be ghost enhancement which 
equates to the Kugo-Ojima criterion being satisfied, \cite{32}. Formally 
writing the radiative corrections to the Faddeev-Popov ghost $2$-point function
as $u(p^2)$ then ghost enhancement follows if $u(0)$~$=$~$-$~$1$ which is the 
Kugo-Ojima condition. This was verified at two loops in the massless quark case
in \cite{16}. Therefore, we have repeated that calculation here with massive 
quarks and examined the zero momentum limit. This involves applying the vacuum 
bubble expansion to the $31$ contributing two loop Feynman diagrams. The 
computation makes use of the master integrals discussed in section $3$ and we 
have used the same routines in order to do the {\sc Form} identifications. The 
outcome is similar to \cite{16}. In other words the Kugo-Ojima criterion is 
satisfied at two loops precisely when the two loop massive quark Gribov gap 
equation is satisfied. Indeed as emphasised in Zwanziger's articles, the theory
has no meaning as a gauge theory unless this occurs. Therefore, we are 
confident that our result (\ref{gap2}) is correct.

\sect{Discussion.}

We conclude with several observations. The inclusion of massive quarks in the
Gribov-Zwanziger approach has not affected the main properties of the 
Faddeev-Popov ghost enhancement at two loops. Moreover, the one loop 
verification of gluon suppression of \cite{16} is also unaffected with massive
quarks. This is because at one loop the diagrams involving massive quarks do
not arise in that part of the matrix of $2$-point functions responsible for the
vanishing of the gluon propagator in the infrared. It is worth noting that our
original expectation was that massive quarks would not upset these key 
properties of the Yang-Mills fields. One of the main outcomes of the result
(\ref{gap2}) is the very much involved form which is clearly due to the
multi-scale nature of the underlying Feynman diagram. Whilst we have 
concentrated on what is now known as the scaling solution, 
\cite{17,18,19,20,21,22}, rather than the decoupling solution it does serve as
an indication of what to expect if one were to study the same problem in the
latter case. For instance, within the Gribov-Zwanziger context, \cite{23,24},
the gluon propagator acquires an additional mass scale deriving from the 
condensation of a mass for the Zwanziger localizing ghost fields. Aside from
giving three scale two loop integrals for Feynman graphs without quarks it will
result in {\em four} scale two loop integrals for the case we studied in depth
here. Clearly that would be a difficult computation. 

\vspace{1cm}
\noindent
{\bf Acknowledgements.} 
F.R. Ford thanks the University of Liverpool for a Research Studentship.


\begin{thebibliography}{99}
\bibitem{1} V.N. Gribov, Nucl. Phys. {\bf B139} (1978), 1. 
\bibitem{2} D. Zwanziger, Nucl. Phys. {\bf B209} (1982), 336. 
\bibitem{3} D. Zwanziger, Nucl. Phys. {\bf B321} (1989), 591. 
\bibitem{4} D. Zwanziger, Nucl. Phys. {\bf B323} (1989), 513. 
\bibitem{5} G. Dell'Antonio \& D. Zwanziger, Nucl. Phys. {\bf B326}
(1989), 333. 
\bibitem{6} G. Dell'Antonio \& D. Zwanziger, Commun. Math. Phys. {\bf 138}
(1991), 291. 
\bibitem{7} D. Zwanziger, Nucl. Phys. {\bf B378} (1992), 525. 
\bibitem{8} D. Zwanziger, Nucl. Phys. {\bf B399} (1993), 477. 
\bibitem{9} D. Zwanziger, Phys. Rev. {\bf D65} (2002), 094039.
\bibitem{10} D. Zwanziger, Phys. Rev. {\bf D69} (2004), 016002.
\bibitem{11} M.A. Semenov-Tyan-Shanskii \& V.A. Franke, Zap. Nauchn. Semin.
LOMI {\bf 120} (1982), 159; J. Sov. Math. {\bf 34} (1986), 1999.
\bibitem{12} P. van Baal, Nucl. Phys. {\bf B369} (1992), 259. 
\bibitem{13} N. Maggiore \& M. Schaden, Phys. Rev. {\bf D50} (1994), 6616.  
\bibitem{14} D. Dudal, R.F. Sobreiro, S.P. Sorella \& H. Verschelde, Phys. Rev.
{\bf D72} (2005), 014016.  
\bibitem{15} J.A. Gracey, Phys. Lett. {\bf B632} (2006), 282.
\bibitem{16} J.A. Gracey, JHEP {\bf 05} (2006), 052.
\bibitem{17} A. Cucchieri \& T. Mendes, PoS LAT2007 (2007), 297.  
\bibitem{18} I.L. Bogolubsky, E.M. Ilgenfritz, M. M\"{u}ller-Preussker \& A. 
Sternbeck, PoS LAT2007 (2007), 290.
\bibitem{19} A. Cucchieri \& T. Mendes, Phys. Rev. Lett. {\bf 100} (2008),
241601. 
\bibitem{20} A. Cucchieri \& T. Mendes, Phys. Rev. {\bf D 78} (2008), 094503. 
\bibitem{21} Ph. Boucaud, J.P. Leroy, A.L. Yaounac, J. Micheli, O. Pene \&
J. Rodriguez-Quintero, JHEP {\bf 06} (2008), 099.
\bibitem{22} A.C. Aguilar, D. Binosi \& J. Papavassiliou, Phys. Rev. {\bf D78}
(2008), 025010. 
\bibitem{23} D. Dudal, S.P. Sorella, N. Vandersickel \& H. Verschelde, Phys.
Rev. {\bf D77} (2008), 071501. 
\bibitem{24} D. Dudal, J.A. Gracey, S.P. Sorella, N. Vandersickel \& H. 
Verschelde, Phys. Rev. {\bf D78} (2008), 065047. 
\bibitem{25} K.G. Chetyrkin, Nucl. Phys. {\bf B710} (2005), 499.
\bibitem{26} P. Nogueira, J. Comput. Phys. {\bf 105} (1993), 279. 
\bibitem{27} J.A.M. Vermaseren, math-ph/0010025.
\bibitem{28} J. van der Bij \& M. Veltman, Nucl. Phys. {\bf B231} (1984), 205.
\bibitem{29} C. Ford, I. Jack \& D.R.T. Jones, Nucl. Phys. {\bf B387} (1992),
373.
\bibitem{30} A.I. Davydychev \& J.B. Tausk, Nucl. Phys. {\bf B397} (1993), 123.
\bibitem{31} L. Lewin, ``Dilogarithms and associated functions'' (Macdonald,
London, 1958).
\bibitem{32} T. Kugo \& I. Ojima, Prog. Theor. Phys. Suppl. {\bf 66} (1979), 1;
Prog. Theor. Phys. Suppl. {\bf 77} (1984), 1121.
\end{thebibliography}
\end{document}